\documentclass[12pt]{article}
\InputIfFileExists{epsf.sty}{}{}
\usepackage{graphicx}
\def\beq{\begin{equation}}
\def\eeq{\end{equation}}
\def\bea{\begin{eqnarray}}
\def\eea{\end{eqnarray}}

\def\roughly#1{\mathrel{\raise.3ex\hbox
{$#1$\kern-.75em\lower1ex\hbox{$\sim$}}}}

\begin{document}

\begin{center}
\bigskip
{\Large \bf Extracting $\beta$ and the new $D_{sJ}$ resonances 
 \footnote{talk given at {\it MRST 2004: From Quarks to
Cosmology}, Concordia University, Montreal, May 2004.}} \\
\bigskip
\bigskip
{\large Alakabha Datta\footnote{datta@physics.utoronto.ca}}

\end{center}

\begin{center}
{ \it Department of Physics, \\
University of Toronto, \\ 
60 St George St,\\
Toronto, Ontario, Canada\\ }
\end{center}

\begin{center} 
\bigskip (\today)
\vskip0.5cm
{\Large Abstract\\}
\vskip3truemm
\parbox[t]{\textwidth} { The three body decays 
$B \rightarrow D^{(*)} \bar{D}^{(*)} K_s$ 
may be used to measure both $\sin{ 2\beta}$
and $\cos{ 2\beta}$. Crucial to the
$\cos{ 2\beta}$ measurement is the resonant contribution to the 
three body decay from p-wave excited $D_s$ states. If these
p-wave states are the newly discovered $D_s(2317)$ and
$D_s(2460)$ then they are below the $D^{(*)} K$ threshold and hence
do not contribute to
$B \rightarrow D^{(*)} \bar{D}^{(*)} K_s$. The three body decays can then 
be used to measure $\sin{ 2\beta}$ without resonant dilution 
and to look for new physics in $ b \to c \bar{c} s $ transition.}
\end{center}

\thispagestyle{empty}
\newpage
\setcounter{page}{1}
\baselineskip=14pt

\section{Introduction}
The decay $B^0 \to J/\psi K_s$ 
provides a clean measurement of the angle $\sin (2\beta)$ in
the unitarity triangle\cite{Sanda}.
Both BaBar and Belle have measured this CP phase, with the world
average being \cite{HFAG}
\begin{equation}
\sin 2\beta = 0.736 \pm 0.049 ~.
\label{sin2beta}
\end{equation}
Other
modes can also provide relevant information on the angle $\beta$,
an example being the decay $B^0 \to D^{(*)} {\overline D}^{(*)}$.
The possibility of extracting $ \cos {2\beta}$ from the decay
$ B^0 \to D {\bar D} K_s $ was mentioned in Ref. \cite{Charles}. 
These modes are enhanced relative to
$ B^0 \to D^{(*)} {\overline D}^{(*)}$  by the 
factor $|V_{cs}/V_{cd}|^2 \sim 20$.
As in the case of $B^0 \to J/\psi K_s$
decay, the penguin contamination is expected to be small in these decays. 
Moreover these decays can be used to probe both $\sin 2 \beta$ and 
$ \cos {2 \beta}$ which can resolve
$\beta \rightarrow \pi/2- \beta$ ambiguity \cite{BDOP}.

\section{ $ \beta$ from $B \rightarrow D^{(*)} \bar{D}^{(*)} K_s$ }
The amplitude for the decay $B^0\to D^* \bar{D}^* K_s$ can  
have a resonant contribution 
and a non-resonant contribution.
 For the resonant contribution
 the $D^* K_s$ in the final state comes dominantly from an
excited $D_s (1^+)$ state.
In the approximation of treating $D^* \bar{D}^* K_s$ as $D^* D_s(excited)$,
there are four possible excited p-wave $D_s$ states which might contribute.
These are the two  states with the light degrees of freedom
in a $j^P=3/2^{+}$ state and the two  states with light
degrees of freedom in a $j^P = 1/2^{+}$ state.
Since the states with $j^P=3/2^{+}$ decay via d-wave to $D^* K_s$, 
they are suppressed. Of the states
with light degrees of freedom in $j^P=1/2^{+}$
states, only the $1^+$ state contributes. The $0^+$ state is forbidden
to decay to the final state $D^* K_s$.

To estimate the above contribution and to calculate the non-resonant
amplitude, we use heavy hadron chiral perturbation theory 
 (HHCHPT)\cite{HHCHPT}.  The momentum $p_k$ of $K_s$ can have a 
maximum value of about  1 GeV for 
$ B^0 \to D^{*+} {\bar D}^{*-} K_s$. This is of the same order as 
$\Lambda_{\chi}$ which sets the scale below which  
we expect HHCHPT to be valid. It follows that in the present case it is 
reasonable to apply HHCHPT to calculate the three body decays. 

In the lowest order in the
HHCHPT expansion, contributions to the decay amplitude come from the 
contact interaction terms and the pole diagrams which give rise to the 
non-resonant and resonant contributions respectively. The pole diagrams get 
contributions from the various 
multiplets involving $D_s$ type resonances as mentioned above. 
In the framework of HHCHPT, the
 ground state heavy meson has the light degrees of freedom in a
spin-parity state $j^P={1\over2}^-$, corresponding to the usual
pseudoscalar-vector meson doublet with $J^P=(0^-,1^-)$.  The first
excited state involves a p-wave excitation, in which the light
degrees of freedom have $j^P={1\over2}^+$ or ${3\over2}^+$.  In the
latter case we have a heavy doublet with $J^P=(1^+,2^+)$. These states can
probably be identified with $D_{s1}(2536)$ and $D_{sJ}(2573)$.
Heavy quark symmetry rules out any pseudoscalar coupling of this
doublet to the ground state at lowest order in 
the chiral expansion \cite{Fluke};
hence the effects of these states will be suppressed and
we will ignore them in our analysis.

The other excited doublet has
$J^P=(0^+,1^+)$.  
These states are expected to decay rapidly through
s-wave pion emission and have large widths \cite{nathanmark}.  
The $1^+$ state in the $D$ system 
has already been seen.
Only the $1^+$ can contribute in this case. For later reference,
we denote this state by $D^{*'}_{s1}$.  
However, quark model estimates suggest \cite{God}
that these states
should have masses near $m+\delta m$ with $\delta m=500$ MeV, 
where $m$ is the mass of the lowest multiplet.

 Assuming that the leading order terms in HHCHPT give 
the dominant contribution to the decay amplitude 
 we will neglect all sub-leading effects 
suppressed by $1/\Lambda_{\chi}$ and
  $1/m$, where m is the heavy quark mass.  It can be  shown that from the 
time dependent analysis of 
 $ B^0(t) \to D^{*+} { D}^{*-}K_s$ one can extract
 $\sin (2\beta)$ and $\cos (2\beta)$.
 Measurement of both
$\sin (2\beta)$ and $\cos (2\beta)$ can resolve the
$\beta \rightarrow \pi/2- \beta$ ambiguity as already mentioned.  
The measurement of
 $\sin (2\beta)$ can be made from the time dependent partial rate asymmetry
while a fit to the time dependent rate for
$\Gamma[ B^0(t) \to D^{*+} { D}^{*-}K_s] +
\Gamma[ {\bar B}^0(t) \to D^{*+} { D}^{*-}K_s] $ may be used for
the extraction of $\cos (2\beta)$.  The
$\cos (2\beta)$ term measures  the overlap of the imaginary part of
the amplitudes for $B \to D^{*+} D^{*-} K_s$ and 
${\bar B} \to D^{*+}D^{*-}K_s$ decays and is non zero only if there is a 
resonance contribution.

As in the case for $ B \to D^{*+} { D}^{*-}$ the asymmetry in
$B \to D^{*+} { D}^{*-} K_s$ is also diluted.
For the non resonant contribution to $B \to D^{*+} { D}^{*-} K_s$ 
 the final state is an admixture of CP states 
with different CP parities.
This leads to the dilution of the asymmetry and this is the same 
dilution of the asymmetry 
as in the case for
$B \to D^{*+} D^{*-}$.
 When the resonant contribution is included there 
is further dilution of the asymmetry from the additional mismatch of
the amplitudes for
 $B$ and ${ \bar B}$ decays. One can reduce the additional dilution
of the CP asymmetry by imposing cuts to remove the resonance. A narrow 
resonance is preferable as it can be more effectively removed 
from the signal region than a broad
resonance . 
When we include  the resonance contribution it turns out that
a broader resonance leads to a larger value of D and is a more
 useful  probe of $\cos (2\beta)$  because
of the 
 the larger overlap of the amplitudes for
$B \to D^{*+} { D}^{*-} K_s$  and 
${\bar B} \to D^{*+} { D}^{*-} K_s$ decays.

The differential decay distribution of the time independent process
$ B^0 \to D^{*+} { D}^{*-}K_s$ can be used  to discover the 
$1^{+}$ resonance $D^{*'}_{s1}$ if it is above the $D^{(*)} K$ threshold.
The differential decay distribution for  
 small values of $E_k$, the kaon energy, 
shows a clear resonant structure which comes from the pole 
contribution to the amplitude  with the excited $J^P=1^+$
 intermediate state. Therefore, 
examination of the $D^{*} K_s$ mass spectrum
may be the best experimental way to find the broad $1^+$ p-wave
$D_s$ meson and  a fit to the
decay distribution will measure its mass and the coupling.
  
The  extraction of $\sin {2 \beta}$ and
$\cos{2\beta}$ from the time dependent rate for
$B(t) \to D^{*+}D^{*-} K_s$ can be done in the following manner:
 We define the following amplitudes
\begin{eqnarray}
{a^{\lambda_1,\lambda_2}}  & \equiv &
 A(B^0(p)\to D^{+*}_{\lambda_1}(p_{+}) D^{-*}_{\lambda_2}(p_{-}) 
K_s(p_k))\\ 
{\bar a^{\lambda_1,\lambda_2}} & \equiv & 
A(\bar B^0(p)\to D^{+*}_{\lambda_1}(p_{+}) D^{-*}_{\lambda_2}(p_{-}) K_s(p_k  ),\label{AAbar}
\end{eqnarray}
where $B^0$ and $\bar  B^0$ represent unmixed neutral $B$ and $\lambda_1$ and $\lambda_2$ are the polarization indices of the $D^{*+}$ and $D^{*-}$ respectively.

The time-dependent amplitudes for an oscillating state $B^0(t)$ which has been tagged as a $B^0$ meson at time $t=0$ is given by, 
\begin{eqnarray}
A^{\lambda_1,\lambda_2}(t)& =& {a}^{\lambda_1,\lambda_2} \cos\left(\frac{\Delta m\,t}2\right) + i e^{-2i \beta}
{\bar{a}}^{\lambda_1,\lambda_2} \sin\left(\frac{\Delta m\,t}2\right)\,,
\end{eqnarray}
and the time-dependent amplitude squared summed over polarizations and integrated over the phase space angles is:
\begin{eqnarray*}
|A(s^+,s^-;t)|^2 & = & \frac{1}{2}\left[ {\rm G}_0(s^+,s^-)+{\rm G}_{\rm c}(s^+,s^-)\cos\Delta m\,t-{\rm G}_{\rm s}(s^+,s^-)\sin\Delta m\,t \right] \, 
\label{osc}
\end{eqnarray*}
with
\begin{eqnarray*}
{\rm G}_0(s^+,s^-)     & = & |{a}(s^+,s^-)|^2 +|\bar{{a}}(s^+,s^-)|^2 ,\\
{\rm G}_{\rm c}(s^+,s^-) & = & |{a}(s^+,s^-)|^2 -|\bar{{a}}(s^+,s^-)|^2 ,\\
{\rm G}_{\rm s}(s^+,s^-) & = & 2\Im\left (e^{-2i \beta} \bar{a}(s^+,s^-)
{{a}^\ast(s^+,s^-)} \right ) \nonumber \\
& = & -2 \sin(2\beta)\, \Re \left ( \bar{a}{{a}^\ast} \right ) + 2\cos (2\beta) \,\Im \left ( \bar{a}{{a}^\ast} \right ).\label{gmix}
\end{eqnarray*}
The variables $s^+$ and $s^-$ are the Dalitz plot variable
$$ s^+= (p_{+} +p_k)^2,\quad s^-= (p_{-} +p_k)^2$$
The transformation defining the CP-conjugate channel 
$\bar B^0(t)\to D^{*-}D^{*+} K_s$ is $s^+\leftrightarrow s^-$, 
${ a}\leftrightarrow \bar{ a}$ and $\beta\to -\beta$. Then:
\begin{eqnarray*}
|\bar A(s^-,s^+;t)|^2 & = &\frac{1}{2}\left[ {\rm G}_0(s^-,s^+)-{\rm G}_{\rm c}(s^-,s^+)\cos\Delta m\,t+{\rm G}_{\rm s}(s^-,s^+)\sin\Delta m\,t \right]\,.
\end{eqnarray*}
Note that for simplicity the $e^{-\Gamma t}$ and constant phase space factors have
been omitted in the above equations.

It is convenient in our case to replace the variables $s^+$ and $s^-$ by 
the variables $y$ and $E_k$ where $E_k$ is the $K_s$ energy in the rest 
frame of the $B$ and $y=\cos{\theta}$ with $\theta$ being the angle between 
the momentum of $K_s$ and $D^{*+}$ in a frame where the two $ D^{*}$ are moving back to back along the z- axis. This frame is boosted with respect to the rest frame of the $B$ with ${\vec {\beta}}=-({\vec{p}_k}/m_B)(1/(1-E_k/m_B))$. Note
$s^+\leftrightarrow s^-$ corresponds to $y \leftrightarrow -y$. The 
variable $y$ can be expressed in terms of variables in the rest frame of $B$. 
For instance
\begin{eqnarray}
E_{+} & = & \frac{E_B^{\prime} E_{+}^{\prime} -{p}_B^{\prime} 
{p}_{+}^{\prime}y}{m_B} \
\end{eqnarray}
where  $E_{+}$ and $E_{+}^{\prime}$ are the energy of the $D^{*+}$ in 
the rest frame of the $B$ and in the boosted frame while $ E_B^{\prime}$ 
is the 
energy of the $B$ in the boosted frame. The magnitudes of the 
momentum of the $B$ and the $D^{*+}$ in the boosted frame are 
given by ${p}_B^{\prime}$ and ${p}_{+}^{\prime}$ respectively.

In the approximation of neglecting the penguin contributions, proportional
to the small CKM elements $V_{ub}V_{us}^*$, to the 
amplitude there is no direct CP violation. This leads to the relation
\begin{equation}
{ a}^{\lambda_1,\lambda_2}(\vec{p}_{k1},E_k)=
{\bar a}^{-\lambda_1,-\lambda_2}(\vec{-p}_{k1},E_k)\
\end{equation}
where $\vec{p}_{k1}$ is the momentum of the of the $K_s$ in the boosted frame.
The above relations then leads to 
\begin{eqnarray}
G_0(-y,E_k) & = & G_0(y,E_k) \\
G_c(-y,E_k) & = & - G_c(y,E_k) \\
G_{s1}(-y,E_k) & = & G_{s1}(y,E_k) \\
G_{s2}(-y,E_k) & = & -G_{s2}(y,E_k) \
\end{eqnarray}
 where we have defined
\begin{eqnarray}
G_{s1}(y,E_k) & = &\Re \left ( \bar{a}{{a}^\ast} \right)   \\
G_{s2}(-y,E_k) & = & \Im \left ( \bar{a}{{a}^\ast} \right)  \
\end{eqnarray}
 Carrying out the integration over the phase space variables $y$ and $E_k$ one 
gets the following expressions for the time-dependent total rates
for $ B^0(t) \to D^{*+}D^{*-}K_s$ and the CP conjugate process
\begin{eqnarray}
\Gamma(t) &= &\frac{1}{2}[I_0 + 2 \sin( 2\beta)\sin (\Delta m t) I_{s1}]\\
{\overline{\Gamma}}(t) &= &\frac{1}{2}[I_0 - 2 \sin( 2\beta)
\sin (\Delta m t) I_{s1}]\
\end{eqnarray}
where $I_0$ and $I_{s1}$ are the integrated $G_0(y,E_k)$ and 
$G_{s1}(y,E_k)$ functions.
One can then extract $\sin (2 \beta)$ from the rate asymmetry
\begin{equation}
\frac{\Gamma(t)-{\overline{\Gamma}}(t)}{\Gamma(t)+{\overline{\Gamma}}(t)} =
D\sin( 2\beta)\sin (\Delta m t)\
\end{equation}
where
\begin{equation} 
 D = \frac{2I_{s1}}{I_0} \
\end{equation}
is the dilution factor. The quantities $I_{s1,0}$ can be calculated in 
HHCHPT\cite{BDOP}.

The $\cos(2 \beta)$ term can be probed by by integrating over half the 
range of the variable $y$ which can be taken for instance to be $y \ge 0$.
In this case we have
\begin{eqnarray*}
{\Gamma}(t) &= &\frac{1}{2}[{J}_0 
+{J}_c \cos (\Delta m t) + 2 \sin( 2\beta)\sin (\Delta m t) {J}_{s1}-
2 \cos( 2\beta)\sin (\Delta m t) {J}_{s2}]\\
{\overline{\Gamma}}(t) &= & \frac{1}{2}[{J}_0 
+{J}_c \cos (\Delta m t) - 2 \sin( 2\beta)\sin (\Delta m t) {J}_{s1}-
2 \cos( 2\beta)\sin (\Delta m t) {J}_{s2}]\
\end{eqnarray*} 
where ${J}_0$, ${J}_c$, ${J}_{s1}$ and  ${J}_{s2}$,  
are the integrated $G_0(y,E_k)$, $G_c(y,E_k)$, $G_{s1}(y,E_k)$ and 
$G_{s2}(y,E_k)$ functions integrated over the range $y \ge 0$.
The quantities $J_{0,c,s1,s2}$ can be calculated in 
HHCHPT\cite{BDOP}.
One can measure $\cos(2 \beta)$  by fitting to the time distribution
of $\Gamma(t) +{ \bar \Gamma}(t)$.
Measurement of the  $\cos(2 \beta)$ then resolves    
the $\beta \to {\pi \over 2} - \beta$ ambiguity.
In passing we note that only the sign of
$\cos(2 \beta)$ is required to  resolve  the   
the $\beta \to {\pi \over 2} - \beta$ ambiguity.
\section{ The new $D_{sJ}$ resonances }
The previous year saw the discovery of an unexpectedly light narrow
resonance in $D^+_s\pi^0$ with a mass of $2317 MeV/c^2$ first reported
 by the BaBar collaboration \cite{babarDs}, together with another second narrow
resonance in $D_s\pi^0\gamma$ with a mass $2460 MeV/c^2$
\cite{cleoDs}.

The smaller than expected masses and narrow widths of these states
have led, among other explanations \cite{jackson}, to a multi-quark
anti-quark or a $D K$ molecule interpretation of these states
\cite{barneslipkin}, or to an interpretation as p-wave states where
the light degrees of freedom are in an angular momentum state $j_q =
{1 \over 2}$ \cite{bardeenhill}, or even some combination of these
\cite{sandip}.  There are also conflicting lattice interpretations of
these states \cite{latticeDs}.  The mass difference between the
$D_s(2317)$ and the well established lightest charm-strange meson,
$D_s$, is $\Delta M = 350 MeV/c^2$. This is less than the kaon mass,
thus kinematically forbidding the decay $D_s(2317) \rightarrow
D_{u,d}+K$.  The possible resonance at $2460 MeV/c^2 $ also has such a
mass difference when taken with the lighter $D^*$ state.  The
interpretation of these states as bound $D^{(*)} K$ molecules just
below the $D^{(*)}K$ threshold is particularly interesting in the
light of the recent discovery of a narrow resonance in the decay
$J/\psi \rightarrow \gamma p \bar{p}$ \cite{bes} which has been
interpreted as a zero baryon number, ``deuteron-like singlet $ {}^1S_0
$'' bound state of $p$ and $\bar{p}$ \cite{datta-baryonium}.

The  production of these new $D_{sJ}$ states in non 
leptonic $B$ decays appear to be in conflict with theory predictions based 
on factorization and heavy quark symmetry \cite{DO,Li}.
Following Ref. \cite{DO}
let us first assume that we can identify the the newly discovered states
$D_s(2317)$ with $D_{s0}$ and $D_s(2460)$ with $D_{s1}^*$.
In the Standard Model (SM) 
the amplitudes for $B \to D^{(*)}D_{s0}(D_{s1}^*)$, 
are generated by the following effective 
Hamiltonian \cite{buras}:
\begin{eqnarray}
H_{eff}^q &=& {G_F \over \protect \sqrt{2}} 
   [V_{fb}V^*_{fq}(c_1O_{1f}^q + c_2 O_{2f}^q) \nonumber\\
& - &
     \sum_{i=3}^{10}(V_{ub}V^*_{uq} c_i^u
+V_{cb}V^*_{cq} c_i^c +V_{tb}V^*_{tq} c_i^t) O_i^q] +H.C.\;,
\end{eqnarray}
where the
superscript $u,\;c,\;t$ indicates the internal quark, $f$ can be $u$ or 
$c$ quark, $q$ can be either a $d$ or a $s$ quark depending on 
whether the decay is a $\Delta S = 0$
or $\Delta S = -1$ process.
The operators $O_i^q$ are defined as \cite{datta-dcp}
\begin{eqnarray}
O_{1f}^q &=& \bar q_\alpha \gamma_\mu Lf_\beta\bar
f_\beta\gamma^\mu Lb_\alpha\;,\;\;\;\;\;\;O_{2f}^q =\bar q
\gamma_\mu L f\bar
f\gamma^\mu L b\;,\nonumber\\
O_{3,5}^q &=&\bar q \gamma_\mu L b
\bar q' \gamma_\mu L(R) q'\;,\;\;\;\;\;\;\;O_{4,6}^q = \bar q_\alpha
\gamma_\mu Lb_\beta
\bar q'_\beta \gamma_\mu L(R) q'_\alpha\;,\\
O_{7,9}^q &=& {3\over 2}\bar q \gamma_\mu L b  e_{q'}\bar q'
\gamma^\mu R(L)q'\;,\;O_{8,10}^q = {3\over 2}\bar q_\alpha
\gamma_\mu L b_\beta
e_{q'}\bar q'_\beta \gamma_\mu R(L) q'_\alpha\;,\nonumber
\end{eqnarray}
where $R(L) = 1 \pm \gamma_5$, 
and $q'$ is summed over all flavors except t.  $O_{1f,2f}$ are the 
current-current operators that represent tree level processes. $O_{3-6}$ are the strong gluon induced
penguin operators, and operators 
$O_{7-10}$ are due to $\gamma$ and Z exchange (electroweak penguins),
and ``box'' diagrams at loop level. The values of the Wilson coefficients
can be found in Ref. \cite{buras}.

In the factorization assumption the amplitude for 
$B \to D^{(*)}D_{s0}(D_{s1}^*)$,  can now be written as
\begin{equation}
M=M_1 +M_2\
\end{equation}
where 
\begin{eqnarray}
M_1 & = &\frac{G_F}{\protect \sqrt{2}} X_1
 <D_{s0}(D_{s1}^*)|\, \bar{s} \gamma_\mu(1-\gamma^5) \, c\, |\, 0>
               <D^{(*)} |\, \bar{c} \, \gamma^\mu (1-\gamma^5) \, b\, |\, B>
\nonumber\\
M_2 & = &\frac{G_F}{\protect \sqrt{2}} X_2
 <D_{s0}(D_{s1}^*)|\, \bar{s} (1+\gamma^5) \, c\, |\, 0>
               <D^{(*)} |\, \bar{c} \, (1-\gamma^5) \, b\, |\, B>  \\
\end{eqnarray}
where
\begin{eqnarray}
        X_1 & = & V_c \left(\frac{c_1}{N_c} + c_2 \right) + 
                  \frac{B_3}{N_c} + B_4
                + \frac{B_9}{N_c} + B_{10} \nonumber\\
        X_2 & = & -2 \, \left(\frac{1}{N_c} B_5 + B_6 + \frac{1}{N_c} B_7 
                + B_8 \right)
\end{eqnarray}
We have defined
\begin{equation}
        B_i = - \sum_{q=u,c,t} c_i^q V_q
\label{nonleptonic}
\end{equation}
with
\begin{equation}
        V_q = V_{qs}^{*} V_{qb}
\end{equation}

In the above equations $N_c$ represents the number of colors.  To
simplify matters we neglect the small penguin contributions and so as
a first approximation we will neglect $M_2$.

We can now define the following ratios
\begin{eqnarray}
R_{D0}= \frac{BR[B \to D D_{s0}]}{BR[B \to D D_s]} \nonumber\\
R_{D^*0}= \frac{BR[B \to D^* D_{s0}]}{BR[B \to D^* D_s]} \nonumber\\
R_{D1}= \frac{BR[B \to D D_{s1}^*]}{BR[B \to D D_s^*]} \nonumber\\
R_{D^*1}= \frac{BR[B \to D^* D_{s1}^*]}{BR[B \to D^* D_s^*]} \
\label{ratios}
\end{eqnarray}
Let us focus on the ratio $R_{D0}$ which within factorization and the 
heavy quark limit can be 
written as
\begin{eqnarray}
R_{D0} & = & |\frac{f_{D_{s0}}}{f_{D_s}}|^2\
\end{eqnarray}
where we have neglected phase space ( and other) effects that are
subleading in the heavy quark expansion.
Similarly we have
\begin{eqnarray}
R_{D1} & = & |\frac{f_{D_{s1}^*}}{f_{D_s^*}}|^2\
\end{eqnarray}

Now in the heavy quark limit $ f_{D_{s0}} = f_{D_{s1}^*}$ and $
f_{D_{s}} = f_{D_{s}^*}$ and so one would predict $R_{D0}  \approx 
R_{D1}$. There have been various estimates of the decay constant
$f_{D_{s0}}$ in quark models and in QCD sum
rule calculations ; these typically find
the p-wave , $j_q= {1 \over 2}$ states to have the similar decay
constants as the ground state mesons. We therefore expect $f_{D_{s0}}
\sim f_{D_s}$ giving in addition to the heavy quark predictions 
\begin{eqnarray}
R_{D0} \approx  R_{D1}  \approx  1 \ 
\label{oldpred}
\end{eqnarray}

Experimentally Belle measures \cite{belletrabelsi}
\begin{eqnarray*} 
BR[B \to D D_s(2317)]BR[D_s(2317) \to D_s \pi^0]
& = & (9.9^{+2.8}_{-2.5} \pm 3.0) \times 10^{-4} \
\label{ds}
\end{eqnarray*}
 The dominant decay of the
$D_s(2317)$ is expected to be through the 
$D_s \pi$ mode \cite{godfreyDs} and so
\begin{eqnarray}
BR[D \to D D_s(2317)] & \approx &  10^{-3}\
\label{ds1}
\end{eqnarray}
 Now using the measured
 branching ratio \cite{PDG}
\begin{eqnarray}
 BR[B^+ \to {\overline{D}}^0 D_s^+] &=& (1.3 \pm 0.4) \times 10^{-2}
\nonumber\\
BR[\overline{B}_d \to D^- D_s^+] &=& (8 \pm 3) \times 10^{-3} \
\end{eqnarray}
 one obtains a combined branching ratio
\begin{eqnarray}
 BR[B \to {D} D_s] & \approx & 10^{-2}\
\end{eqnarray}
This leads to $R_{D0} \approx { 1 \over 10}$ (or,$ f_{D_{s0}} \sim {1
  \over 3} f_{D_s}$) which is a factor 10 smaller then theoretical
expectations.

One might argue that factorization is not applicable to 
$ B \to D^{(*)}D^{(*)}$ decays. However recent analysis in 
Ref. \cite{luorosner} find that factorization works well for these decays.
Moreover the quantities in  Eq. \ref{ratios} are ratios of nonleptonic
decay amplitudes and so nonfactorizable effects may cancel. So what one really
requires is significantly different nonfactorizable corrections
between decays with the p-wave states in the final state and 
decays with the ground state mesons in the final state.
It is possible that the discrepancies between experiments and theory may 
arise from a combination of incorrect model prediction of p-wave 
state properties and nonfactorizable effects.

If these new $D_{sJ}$ states are indeed the p-wave $ c \bar{s}$ meson then
they cannot contribute to
$B \rightarrow D^{(*)} \bar{D}^{(*)} K_s$  as they are
 below the $D^{(*)} K$ threshold. Hence to the extent 
that the corrections to predictions of HHCHPT are small the three body decays
$B \rightarrow D^{(*)} \bar{D}^{(*)} K_s$  should be dominated by the 
non resonant contribution. This would then imply that 
$\sin {2 \beta}$ can be cleanly measured in this decay. On the other hand
if the new $D_{sJ}$ states are not the p-wave $ c \bar{s}$ mesons 
but something exotic like four quark states or molecules then experimentally the real 
p-wave states should show up.
   
\section{New Physics}
There  are many reason to believe that the Standard Model is not a complete theory as it leaves several puzzles unresolved, specially in the flavour sector.
There are several hints of possible deviations from the SM 
\cite{Datta,DattaLon}
and several interesting methods have been proposed to measure the parameters
of the underlying new physics \cite{DattaLon}. If there is 
new physics in $ b \to s $ transition then this should show up in decays
with the underlying quark transition $ b \to c \bar{c} s$ which can 
then be probed in $B \rightarrow D^{(*)} \bar{D}^{(*)} K_s$.
One can probe for this new physics also in other decays with the same quark level transition such as $ B \to J/\psi K_s$. However, it is possible that the matrix element of the new physics may be suppressed in some decays and 
enhanced in other making it mandatory to study as many different 
decays as possible. If the new physics involves a new effective
$ b \to s g$ vertex its effect in $ B \to J/\psi K_s$ may be negligible 
because of OZI suppression. Such a suppression would not apply to
 $B \rightarrow D^{(*)} \bar{D}^{(*)} K_s$ decays.

\section{Conclusion} In this talk we have pointed out that
the three body decays 
$B \rightarrow D^{(*)} \bar{D}^{(*)} K_s$ 
may be used to measure both $\sin{ 2\beta}$
and $\cos{ 2\beta}$. The
$\cos{ 2\beta}$ measurement requires resonant contribution to the 
three body decay from p-wave excited $D_s$ states. 
We discussed the newly discovered $D_{sJ}$ states and pointed out that
 their properties are sometimes inconsistent with  an interpretation
 as p-wave $ c \bar {s}$ states.
If these
p-wave states are the newly discovered $D_s(2317)$ and
$D_s(2460)$, , then they are below the $D^{(*)} K$ threshold and hence
do not contribute to
$B \rightarrow D^{(*)} \bar{D}^{(*)} K_s$. The three body decays can then 
be used to measure $\sin{ 2\beta}$ without resonant dilution 
and to look for new physics in $ b \to c \bar{c} s $ transition.

\section*{Acknowledgments}
This work is supported by grant from NSERC. I also thank my collaborators
Pat O'Donnell, S. Pakvasa, T. Browder and D. London for this work.

\end{document}